\documentclass[prb,twocolumn,amsmath,amssymb,superscriptaddress,aps]{revtex4-1}

\usepackage[pdftex]{graphicx}
\usepackage{dcolumn}
\usepackage{amssymb} 
\usepackage{placeins}
\usepackage{tikz}
\usetikzlibrary{shapes,arrows}
\usepackage{amsmath}
\usepackage[utf8]{inputenc}
\usepackage{mathtools}

\usepackage{bm}
\usepackage{hyperref} 
\graphicspath{{figures/}}

\begin{document}

\title{High throughput computational screening for 2D ferromagnetic materials: the critical role of anisotropy and local correlations} 

\author{Daniele Torelli}
\affiliation{Computational Atomic-scale Materials Design (CAMD), Department of Physics, Technical University of Denmark, DK-2800 Kgs. Lyngby, Denmark}
\author{Kristian S. Thygesen}
\affiliation{Computational Atomic-scale Materials Design (CAMD), Department of Physics, Technical University of Denmark, DK-2800 Kgs. Lyngby, Denmark}
\affiliation{Center for Nanostructured Graphene (CNG), Department of Physics, Technical University of Denmark, DK-2800 Kongens Lyngby, Denmark}
\author{Thomas Olsen}
\affiliation{Computational Atomic-scale Materials Design (CAMD), Department of Physics, Technical University of Denmark, DK-2800 Kgs. Lyngby, Denmark}

\date{\today}

\begin{abstract}
The recent observation of ferromagnetic order in two-dimensional (2D) materials has initiated a booming interest in the subject of 2D magnetism. In contrast to bulk materials, 2D materials can only exhibit magnetic order in the presence of magnetic anisotropy. In the present work we have used the Computational 2D Materials Database (C2DB) to search for new ferromagnetic 2D materials using the spinwave gap as a simple descriptor that accounts for the role of magnetic anisotropy. In addition to known compounds we find 12 novel insulating materials that exhibit magnetic order at finite temperatures. For these we evaluate the critical temperatures from classical Monte Carlo simulations of a Heisenberg model with exchange and anisotropy parameters obtained from first principles. Starting from 150 stable ferromagnetic 2D materials we find five candidates that are predicted to have critical temperatures exceeding that of CrI$_3$. We also study the effect of Hubbard corrections in the framework of DFT+U and find that the value of U can have a crucial influence on the prediction of magnetic properties. Our work provides new insight into 2D magnetism and identifies a new set of promising monolayers for experimental investigation.
\end{abstract}

\pacs{Valid PACS appear here}
\maketitle

\section{Introduction}
The nature of magnetic order in two-dimensional (2D) materials is fundamentally different from the three-dimensional case. In 3D, magnetic order arises from spontaneously broken symmetry of the magnetization direction and the magnetic anisotropy only plays a marginal role. In 2D, however, the Mermin-Wagner theorem\cite{Studio2005} prohibits a broken symmetry phase at finite temperatures and the spin rotational symmetry has to be broken explicitly by magnetic anisotropy.

In 2017, two examples of 2D ferromagnetic insulators were discovered experimentally. 1) A monolayer of CrI$_3$ that exhibits magnetic order below 45 K.\cite{HuangClark2017} 2) A bilayer of Cr$_2$Ge$_2$Te$_6$ with a Curie temperature of 20 K.\cite{Gong2017b} In the case of CrI$_3$ the the magnetic order is driven by a strong out-of-plane magnetic anisotropy - a case that is often referred to as Ising-type ferromagnet. In contrast, Cr$_2$Ge$_2$Te$_6$ has a rather weak magnetic anisotropy and the magnetic order is maintained in bilayer structures by interlayer exchange couplings. Subsequently, ferromagnetic order at room temperature has been reported in monolayers of MnSe$_2$\cite{OHara2018} and VSe$_2$.\cite{Bonilla2018} Both of these are itinerant (metallic) ferromagnets and the origin of magnetism in these materials is still not completely clarified. In particular, VSe$_2$ has an easy plane, which implies lack of magnetic order by virtue of the Mermin-Wagner theorem. However, such a two-dimensional spin system may comprise an example of a Kosterlitz-Thouless phase,\cite{Thouless1973} which is known to display magnetic order due to finite size effects.\cite{Holdsworth1994} More recently, Fe$_3$GeTe$_2$\cite{Fei2018} was reported to host itinerant ferromagnetic order below 130 K, which originates from strong out-of-plane magnetic anisotropy. Several other 2D materials have been predicted to exhibit either ferromagnetic or anti-ferromagnetic order based on first principles calculations, but in most cases the predictions have not yet been confirmed by experiments\cite{Gong2019} and estimates of the critical temperatures are often unjustified or very crude.

Two-dimensional CrI$_3$ has proven to comprise a highly versatile material. For example, an applied electric field can induce Dzyaloshinskii-Moriya interactions,\cite{Liu2018a} and switch the magnetic state in bilayer samples.\cite{Jiang2018,Huang2018a} In addition, it has been demonstrated that one can obtain control of in-plane conductivity and valley polarization by constructing heterostructures of CrI$_3$/graphene\cite{Cardoso2018} and CrI$_3$/WSe$_2$\cite{Taniguchi2017} respectively. van der Waals heterostructures of 2D materials involving magnetic layers thus constitute a highly flexible platform for designing spin tunnel junctions and could provide new ways to build nanostructured spintronics devices.\cite{Burch2018, Gong2019, Wang2018a} However, in order to make 2D magnetism technologically relevant there is a pressing need to find new 2D materials that exhibit magnetic order at higher temperatures.

A theoretical search for materials with particular properties may be based on either experimental databases such as the Inorganic Crystal Structure Database (ICSD) or computational databases where first principles simulations are employed to predict new stable materials. The former approach has been applied to predict new 2D materials rooted in exfoliation energies of 3D materials in the ICSD\cite{Bjorkman2012, Ashton2017, Mounet2018} and several materials was found to have a magnetically ordered ground state (at $T=0$ K). An example of the latter approach is the Computational 2D Materials Database (C2DB);\cite{Haastrup2018, Olsen} presently containing 3712 2D materials of which 20 {\%} are predicted to be stable. One advantage of using a theoretical database is that the search is not restricted to materials that are experimentally known in a 3D parent van der Waals structure. However, materials predicted from theoretical databases may pose severe challenges with respect to synthetization and experimental characterization; even if they are predicted to be stable by first principles calculations.

Regarding the magnetic properties of materials, a major difficulty stems from the fact that standard first principles methods can only predict whether or not magnetic order is present at $T=0$ K. For 2D materials the Mermin-Wagner theorem implies that magnetic order at $T=0$ vanishes at any finite temperature in the absence of magnetic anisotropy. A first principles prediction of magnetic order at $T=0$ is therefore irrelevant unless other properties of the material are taken into account. The question then arises: how to calculate the critical temperature for magnetic order given a set of exchange and anisotropy parameters for a particular material. It is clear that the Mermin-Wagner theorem disqualifies any standard mean-field approach because such methods neglect the fluctuations that are responsible for deteriorating magnetic order at finite temperatures in the absence of magnetic anisotropy. On the other hand, the importance of having magnetic anisotropy and an easy axis for the magnetization (as opposed to an easy plane) has led many authors to derive the magnetic properties from an Ising model for which the critical temperatures are known for all Archimedian lattices.\cite{Malarz2005, Kan2014} However, the Ising model only provides a good magnetic model in the limit of infinite single-ion anisotropy and simply provides an upper bound for the critical temperature in general.\cite{Torelli2018} For example, in the case of CrI$_3$, which is regarded as an Ising-type ferromagnet, the Ising model overestimates the critical temperature by a factor of three. The effect of finite anisotropy was analyzed in Ref. \onlinecite{Torelli2018}, where Monte Carlo simulations and renormalized spin-wave theory were applied to obtain a simple expression for the critical temperature of Ising-type ferromagnets. The expression only depends on the number of nearest neighbors, the nearest neighbor exchange interactions, and two anisotropy parameters. In the present work we have applied this expression to search the C2DB for ferromagnetic materials with finite critical temperatures. For some materials in the C2DB, the magnetic structure is not well approximated by an Ising-type ferromagnet and we have performed full Monte Carlo simulations to obtain the critical temperatures of these materials.

The paper is organized as follows. In Sec. \ref{sec:methods} we summarize the computational details and discuss the Heisenberg model, which forms the basis for calculations of critical temperatures in the present work. In Sec. \ref{sec:results} we present the magnetic materials found by searching the C2DB and discuss and compare the calculated critical temperatures with previous works. Sec. \ref{sec:conclusions} contains a conclusion and outlook.

\section{Method and Computational Details}\label{sec:methods}
The materials in C2DB have been found by performing first principles calculations of hypothetical 2D materials in the framework of density functional theory (DFT) with the Perdew-Burke-Ernzerhof (PBE) exchange-correlation functional using the electronic structure package GPAW.\cite{Enkovaara2010a,Larsen2017} The geometry of all materials are fully optimized and the dynamical stability is obtained based on phonon frequencies at the the center and corners of the Brillouin zone. The heat of formation is calculated with respect to standard references\cite{Haastrup2018} and a material is regarded as thermodynamically stable if it situated less than 0.2 eV above the convex hull defined by the 2807 most stable binary bulk compounds from the OQMD.\cite{Saal2013} We find that more than 700 materials in the C2DB are predicted to have a ferromagnetic ground state and $\sim$150 of these are thermodynamically and dynamically stable. The DFT calculations show whether or not the materials have a ferromagnetic ground state at $T=0$ and for insulators the critical temperature can then be obtained from the descriptor derived in Ref. \onlinecite{Torelli2018} or Monte Carlo simulations. The procedure requires knowledge of exchange and anisotropy parameters, which can be obtained from an energy mapping analysis including spin-orbit coupling non-selfconsistently.\cite{Olsen2016a} We will briefly outline the approach below.

The magnetic properties of a system of localized spins are commonly analyzed in terms of the Heisenberg model. The most basic ingredient in the model is the isotropic exchange interactions arising between neighboring spins as a consequence of Coulomb repulsion and Pauli exclusion. In addition, spin-orbit coupling may lead to magnetic anisotropy, which manifests itself through anisotropic exchange interactions\cite{Xu2018} as well as single-ion anisotropy. 2D materials often exhibit (nearly) isotropic magnetic interactions in the plane of the materials and in the following we will restrict ourselves to models of the form
\begin{equation}
\label{eq:heis}
    H = - \frac{1}{2}\sum_{i\neq j} J_{ij} \mathbf{S}_i\cdot\mathbf{S}_j -\sum_{i}A_i(S_i^z)^2- \frac{1}{2}\sum_{i\neq j}B_{ij}S_i^zS_j^z,
\end{equation}
where the sums over $i$ and $j$ run over all magnetic sites. $J_{ij}$ denotes the isotropic exchange between spins at site $i$ and $j$, $B_{ij}$ is the anisotropic exchange for spins pointing out of the plane (here assumed to be the $z$-direction), and $A_i$ is the single-ion anisotropy. We will also assume that the model is composed of a single kind of magnetic atom, which is characterized by a half-integer $S$, yielding the maximum possible eigenvalue of $S_z$ for any site. The most general form of the exchange interaction between sites $i$ and $j$ can be written as $\sum_{\alpha\beta}S_i^\alpha J^{\alpha\beta}_{ij}S_j^\beta$, where $J^{\alpha\beta}_{ij}$ is a $3\times3$ tensor for a given pair of $i$ and $j$. This includes the Dzyaloshinkii-Moriya interactions as the anti-symmetric part as well as symmetric off-diagonal components that give rise to Kitaev interactions.\cite{Xu2018} Such terms are neglected in the present work since, we are mainly interested in critical temperatures which is dominated by the terms included in Eq. \eqref{eq:heis}. However, we emphasize that the neglect of such terms as well as the assumption of in-plane magnetic isotropy is an approximation we will make to reduce the set of parameters needed for the identification of promising candidates. Below we discuss a few important exceptions exemplified by materials with large critical temperatures that are not well described by the model \eqref{eq:heis}.

\begin{figure}[tb]
	\includegraphics[width = 0.5\textwidth]{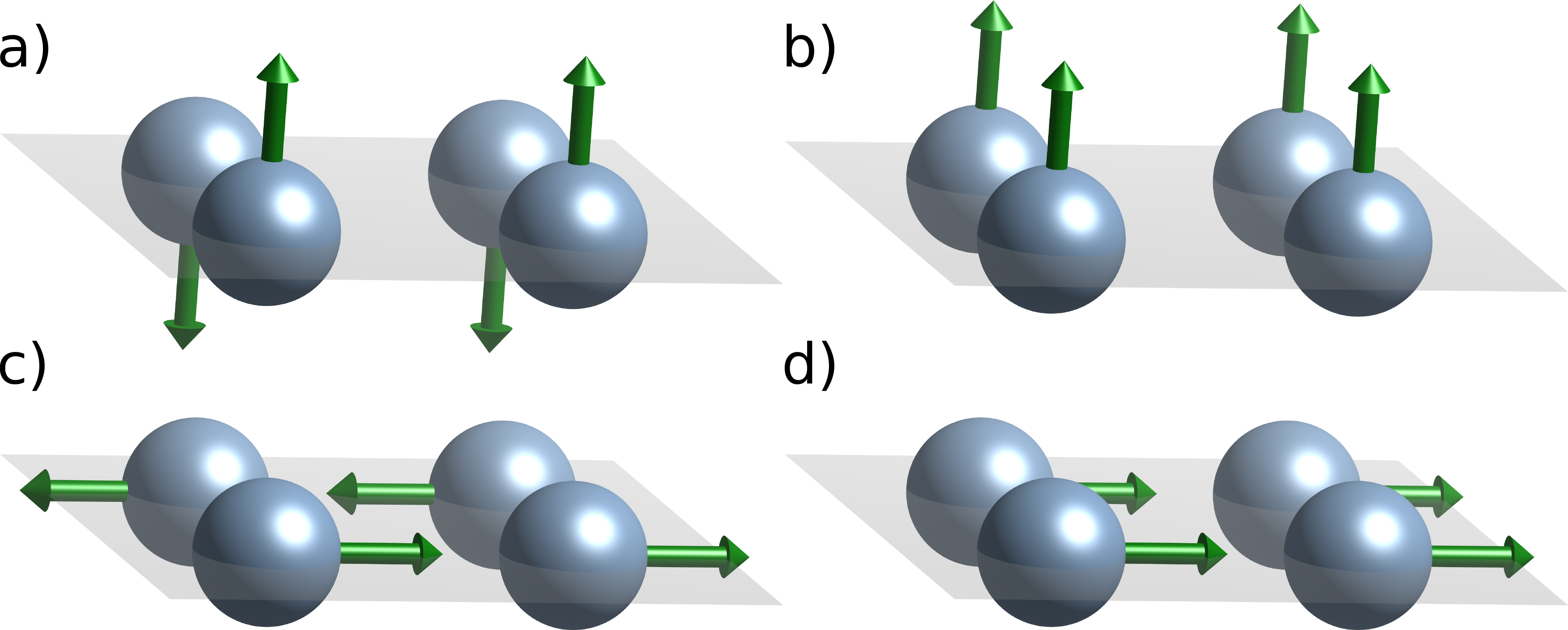}
	\caption{Examples of spin configurations for the calculation of Heisenberg parameters $A$, $B$ and $J$: (a) $E_{\mathrm{\mathrm{AFM}}}^{\perp}$, (b) $E_{\mathrm{FM}}^{\perp}$, (c) $E_{\mathrm{AFM}}^{\parallel}$ and (d) $E_{\mathrm{FM}}^{\parallel}$.   \label{fig:spin_confi}}
\end{figure}
In order to obtain the magnetic properties of a given material based on the model \eqref{eq:heis}, one needs to extract the parameters $J_{ij}$, $A_i$ and $B_{ij}$. In the case of a single magnetic element we have $A_i=A$ and restricting ourselves to nearest neighbor interactions we take $J_{ij}=J$ and $B_{ij}=B$ if $i,j$ are nearest neighbors and $J_{ij}=B_{ij}=0$ otherwise. The parameters can then be obtained by mapping the model to first principles calculations based on density functional theory.\cite{Olsen2017} In particular, the three parameters can be obtained from the total energies of the four spin configurations $E_{\mathrm{FM}}^{\perp(\parallel)}$ and $E_{\mathrm{AFM}}^{\perp(\parallel)}$, where $E_{\mathrm{FM}}$ is the energy of a fully ferromagnetic configuration and $E_{\mathrm{AFM}}$ is an anti-ferromagnetic state that involves anti-parallel spin alignment. The superscripts $\perp$ and $\parallel$ indicates whether the spinors are lying in the plane of the materials or perpendicular to the plane. A ferromagnetic material with $E_{\mathrm{FM}}^{\perp}<E_{\mathrm{FM}}^{\parallel}$, will thus have an out-of-plane easy axis. The four configurations are illustrated in Fig. \ref{fig:spin_confi}. All energies are evaluated with the geometry obtained from the relaxed ferromagnetic ground state. For materials with a single magnetic atom in the unit cell, we have doubled the unit cell in order to accommodate the anti-ferromagnetic configuration. Comparing with Eq. \eqref{eq:heis} and approximating the spin operators by classical vectors we can obtain the parameters as
\begin{align}
 A &=  \frac{\Delta E_{\mathrm{FM}}(1-\frac{N_{\mathrm{FM}}}{N_{\mathrm{AFM}}})+\Delta E_{\mathrm{AFM}}(1+\frac{N_{\mathrm{FM}}}{N_{\mathrm{AFM}}})}{2S^2},\label{eq:A}\\
 B &=  \frac{\Delta E_{\mathrm{FM}}-\Delta E_{\mathrm{AFM}}}{N_{\mathrm{AFM}}S^2},\label{eq:B}\\
 J &= \frac{ E_{A\mathrm{FM}}^{\parallel}- E_{\mathrm{FM}}^{\parallel}}{N_{\mathrm{AFM}}S^2},\label{eq:J}
\end{align}
where $\Delta E_{\mathrm{FM}(\mathrm{AFM})}=E_{\mathrm{FM}(\mathrm{AFM})}^{\parallel}-E_{\mathrm{FM}(\mathrm{AFM})}^{\perp}$ are the energy differences between in-plane and out-of-plane spin configurations for ferromagnetic(anti-ferromagnetic) structures and $N_{\mathrm{FM}(\mathrm{AFM})}$ is the number of nearest neighbors with aligned(anti-aligned) spins in the anti-ferromagnetic configuration. For bipartite lattices $N_{\mathrm{FM}}=0$ and $N_{\mathrm{AFM}}$ is simply the number of nearest neighbors, but the triangular magnetic lattices (for example the MoS$_2$ crystal structure) have no natural anti-ferromagnetic configurations and one has to consider a frustrated configuration where each atom has aligned as well as anti-aligned nearest neighbors.

\begin{figure}[tb]
	\includegraphics[width = 0.5\textwidth]{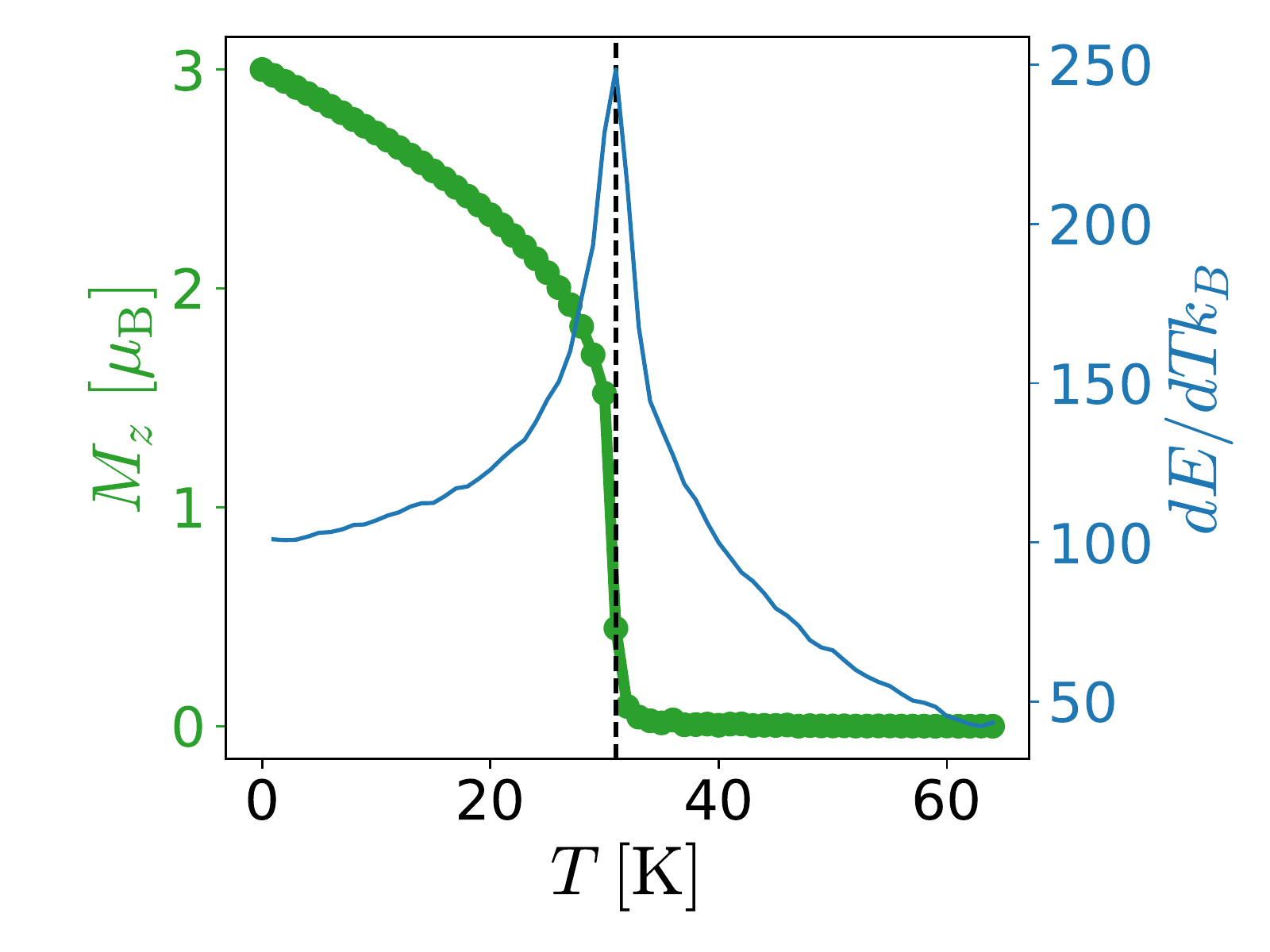}
	\caption{MC calculations of the magnetic moment per atom and heat capacity ($dE/dT$) calculated as a function of temperature for CrI$_3$. The dashed vertical line at $T=31$ K, indicates the predicted critical temperature obtained from Eq. \eqref{eq:tc}.\label{fig:mc}}
\end{figure}

Once the parameters have been determined the Curie temperatures can be calculated from an expression obtained in Ref. \onlinecite{Torelli2018}, which was derived by fitting to classical Monte Carlo (MC) simulations of the model \eqref{eq:heis} combined with renormalized spinwave theory. This simplifies the procedure significantly compared to running MC calculations for all materials and still assures a good reliability compared to available experimental data.\cite{Torelli2018} The expression for the Curie temperature is.
\begin{equation}
\label{eq:tc}
T_\mathrm{C} = \frac{S^2 J T_\mathrm{C}^{\mathrm{Ising}}}{k_\mathrm{B}}f\left( \frac{\Delta}{J(2S-1)} \right )
\end{equation}
where 
\begin{equation}
f(x) = \tanh^{1/4} \left [  \frac{6}{N_{nn}} \log(1+\gamma x) \right]
\end{equation}
is a fitted function with $\gamma = 0.033$ and $N_{nn}$ the number of nearest neighbors. $T_\mathrm{C}^{\mathrm{Ising}}$ is the critical temperature of the corresponding Ising model (in units of $JS^2/k_\mathrm{B}$), which has been tabulated for all the Archimedian lattices,\cite{Malarz2005}  and are  given by 1.52, 2.27, and 3.64 for honeycomb, square and quadratic lattices respectively.
\begin{equation}
\label{eq:delta}
\Delta = A(2S-1)+BSN_{nn}    
\end{equation}
is the spinwave gap derived from spinwave theory. It follows from the Mermin-Wagner theorem that a positive spinwave gap is a minimal requirement for magnetic order in 2D, and $\Delta$ thus provides a crucial parameter that can be used for a rough screening for materials that exhibit magnetic order at finite temperature. It should be noted that for $S=1/2$, the single-ion anisotropy alone cannot open a gap in the spinwave spectrum and magnetic order thus requires anisotropic exchange. We note that the present approach can lead to situations where a material with an out-of-plane easy axis ($\Delta E_{\mathrm{FM}}>0$) has a negative spinwave gap indicating that the ground state is unstable. For example, for a honeycomb lattice with $N_{nn}=N_{\mathrm{AFM}}=3$ and $S=1$ Eqs. \eqref{eq:A}-\eqref{eq:J} one obtains $\Delta<0$ if $\Delta E_{\mathrm{AFM}}>3\Delta E_{\mathrm{FM}}$. This is due to the factor of $2S-1$, which replaces a factor of $2S$ when quantum corrections to the anisotropy terms are taken into account in \textit{renormalized} spinwave theory.\cite{Torelli2018, Gong2017b} In principle this is inconsistent with the energy mapping approach, which is based on a classical treatment of the Heisenberg model. However, a full quantum mechanical energy mapping analysis is beyond the scope of the present work. In Fig. \ref{fig:mc} we compare the magnetization and heat capacity obtained from MC calculations of CrI$_3$ as well as the model result from Eq. \eqref{eq:tc}. The critical temperature can be obtained from the position of the peak in the heat capacity.

\begin{figure}[tb]
	\includegraphics[width = 0.5\textwidth]{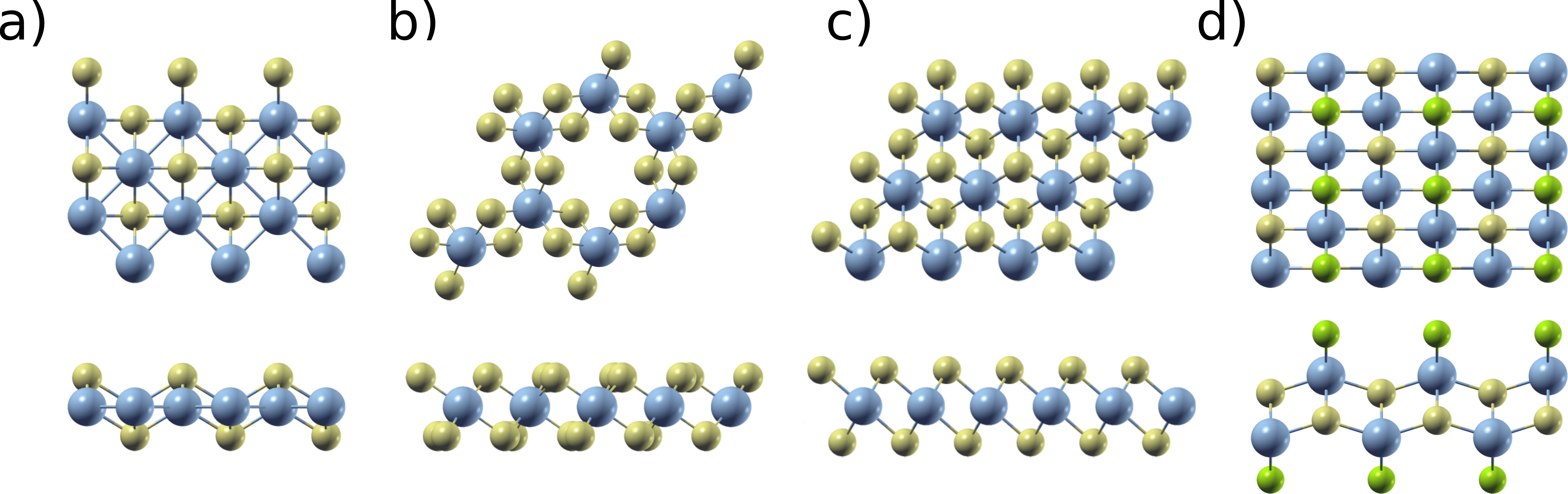}
	\caption{Top and side view of a (a) square, (b) honeycomb, (c) triangular, (d) TMHC crystal structures. Magnetic atoms are in blue.  \label{fig:example_structures}}
\end{figure}
\begin{figure}[tb]
	\includegraphics[width = 0.5\textwidth]{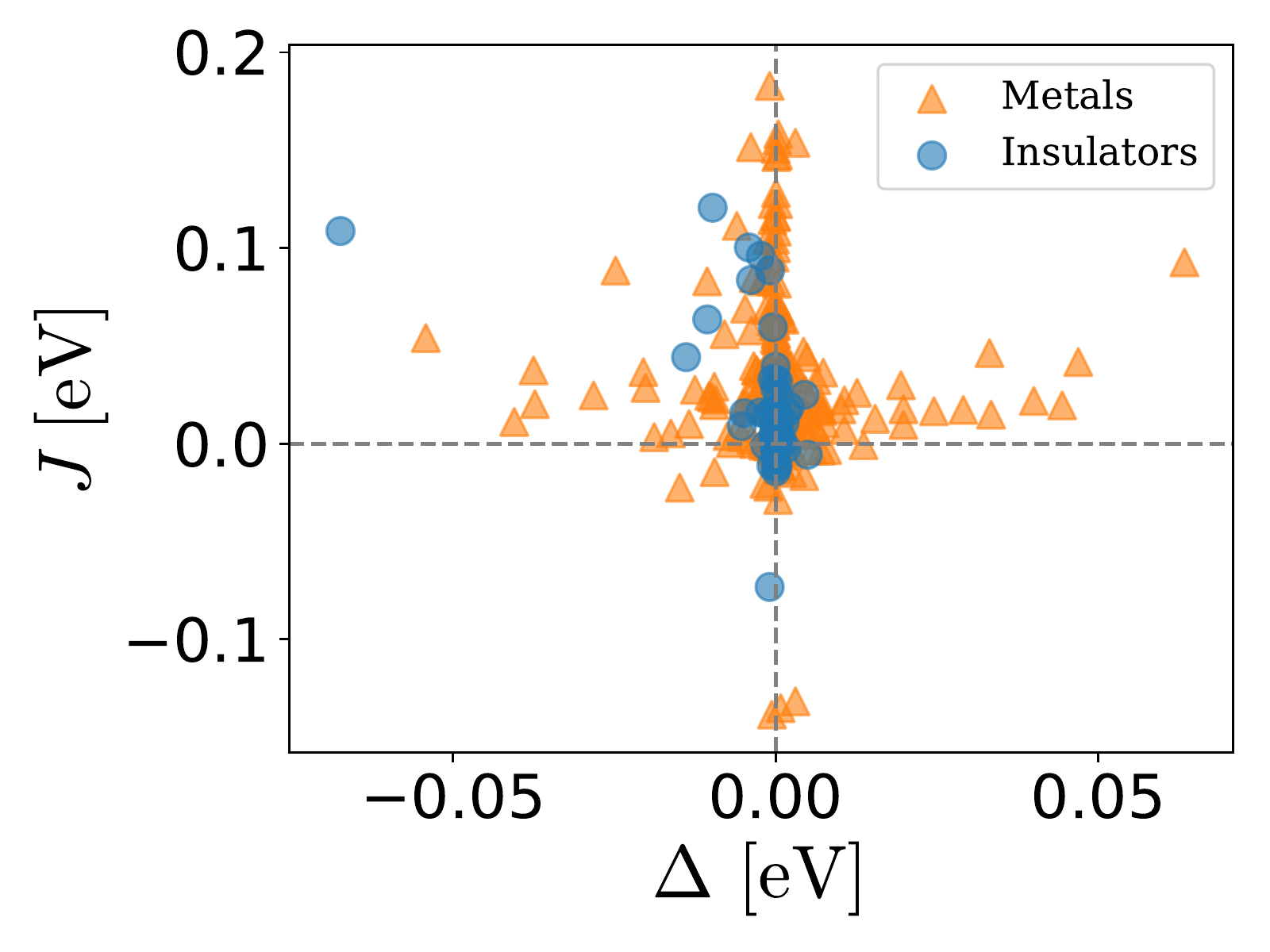}
	\caption{Distribution of the calculated parameters $J$ and $\Delta$ for 87 metallic and 270 insulating materials obtained with PBE.}\label{fig:delta_j}
\end{figure}

The parameters $A$, $B$, and $J$ (Eqs. \eqref{eq:A}-\eqref{eq:J}) and critical temperatures (Eqs. \eqref{eq:tc}-\eqref{eq:delta}) have been calculated for the nearly 550 materials listed in the C2DB database, which display honeycomb, square or triangular magnetic lattices, including stable as well as unstable materials. The calculations were performed with the same plane wave cutoff and $k$-point sampling as used for the magnetic anisotropy calculations in the database.\cite{Haastrup2018} Examples of such structures are shown in Fig. \ref{fig:example_structures} and includes the transition metal dichalcogenides (TMD) in the 1T phase and in the 2H phase (triangular magnetic lattice), compounds adopting the FeSe crystal structure (square magnetic lattice), and transition metal trihalides such as CrI$_3$ (honeycomb magnetic lattice). In Fig. \ref{fig:delta_j} we show all the calculated parameters $J$ and $\Delta$ for insulators and metals with triangular, square or honeycomb lattice. The spinwave gap was calculated by taking the ground state to have an out-of-plane ferromagnetic magnetization and a negative spinwave gap thus implies that the ground state must have in-plane magnetization.

The transition metal halogen chalcogen (TMHC) comprises another crystal structure that deserves an additional comment here. These materials display an atomic structure that resembles a distorted hexagonal magnetic lattice arranged over two layers. Although at least two comparable - but distinct - exchange paths are identifiable, MC calculations show that we can obtain rough estimates of the critical temperatures from the model \eqref{eq:tc} by treating it as an hexagonal lattice with a single effective nearest neighbour coupling obtained from the energy mapping analysis. For example, for CrIS we obtain $T_\mathrm{C}=118$ K from the model \eqref{eq:tc}, which is in decent agreement with the MC results of 140 K including both nearest and next-nearest neighbour exchange interactions.

\section{Results and Discussion}\label{sec:results}
\begin{figure*}[tb]
	\includegraphics[width=\textwidth]{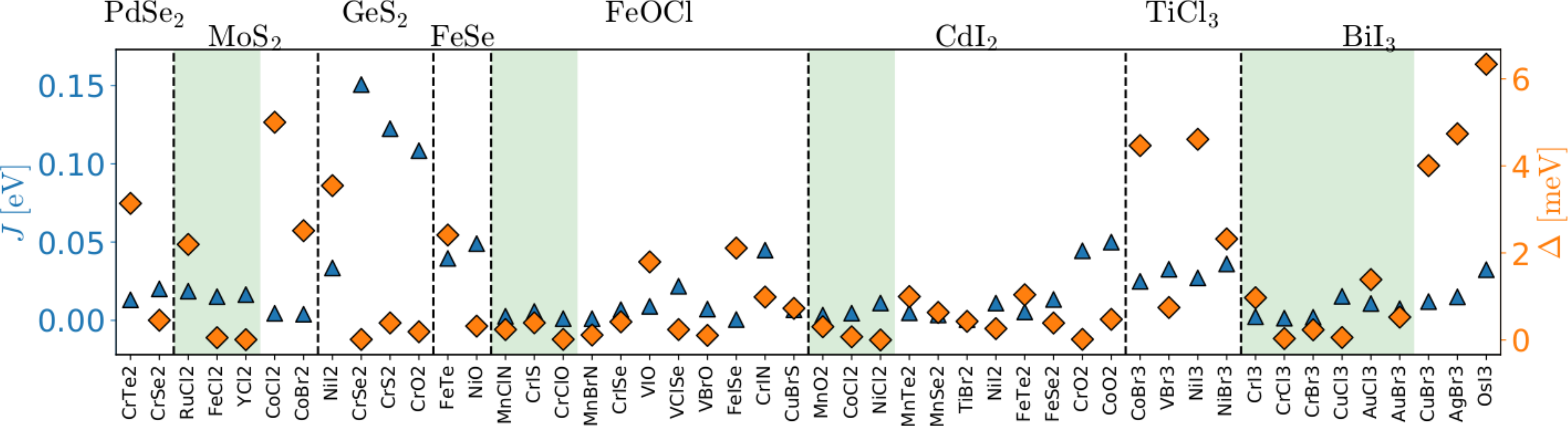}
	\caption{PBE calculations of exchange coupling $J$ (triangles) and spinwave gap $\Delta$ (squares) of stable ferromagnetic materials with $\Delta > 0$. Green background indicates insulating materials. \label{fig:tc_nou}}
\end{figure*}
In Fig. \ref{fig:tc_nou} we show the exchange coupling $J$ and spinwave gap $\Delta$ for the stable ferromagnetic materials with $\Delta>0$. We have performed the calculations for insulating as well as metallic materials. For metals, the value of $S$ is ill-defined and here we have simply used the magnetic moment localized on the magnetic atoms, which is obtained by integrating the magnetization density over the PAW spheres. Moreover, the Heisenberg model is not a reliable starting point for itinerant magnets and to our knowledge there is no simple method to obtain critical temperatures for metallic ferromagnetic materials. For this reason we will not discuss metallic materials any further in the present work, but simply note that large anisotropies and exchange couplings indicate that metallic compounds such as CoBr$_3$,  VBr$_3$, NiI$_3$, and NiBr$_3$ could potentially exhibit very high critical temperatures. In general we observe that most compounds contain transition metal atoms with 3$d$ valence electrons. In particular Cr, Mn, Fe, Ni and Co, which are all well-known elements in magnetic materials. In addition, most of the compounds contain halides, albeit with a few important exceptions (for example MnO$_2$).

In Tab. \ref{tab:no_u} we show a list of the all the insulating ferromagnetic materials and the calculated critical temperatures. The top part of the table contains the stable materials and in the lower part we have included a few examples of materials that exhibit very high critical temperatures but which are predicted to be unstable in their pristine form.
\begin{table}[tb]
  \begin{center}
\begin{tabular}{lccccc}
  Formula       & Structure & $J$ [meV] & $\Delta$ [meV] & $S$ [$\hbar$] & $T_\mathrm{C}$ [K]           \\
    \hline
    \hline
FeCl$_2$  &   MoS$_2$  &   15.2 &  0.056  &  2.0  &  208 \\
CuCl$_3$  &   BiI$_3$  &   15.3  &  0.058  &  1.0  &  37  \\
CrI$_3$  &   BiI$_3$  &   2.3  &  0.96  &  1.5  &  35  \\
CoCl$_2$  &   CdI$_2$  &   2.0 &  0.058  &  1.5  &  31 \\
CrBr$_3$  &   BiI$_3$  &   2.0  &  0.23  &  1.5  &  23  \\
MnO$_2$  &   CdI$_2$  &   0.54  &  0.31  &  1.5  &  19 \\
NiCl$_2$  &   CdI$_2$  &   7.2 &  0.001  &  1.0  &  14 \\
CrCl$_3$  &   BiI$_3$  &   1.4  &  0.033  &  1.5  &  13  \\
\hline
RuCl$_2$  &   MoS$_2$  &   18.7 &  2.3 &    2.0  &  606 \\ 
RuBr$_2$  &   MoS$_2$  &   16.1 &  1.77    &  2.0  &  509
\end{tabular}
\end{center}
\caption{List of 2D magnetic insulating materials with positive exchange coupling $J$ and positive spinwave gap $\Delta$. Structure denotes the prototypical crystal structure and $S$ is the spin carried by each magnetic atom. The critical temperature $T_\mathrm{C}$ is obtained from Eq. \eqref{eq:tc}. The top part of the table contains dynamically and thermodynamically stable materials. The lower part of the table contains materials that are not expected to be stable in their pristine form but exhibit high critical temperatures.
\label{tab:no_u}}
\end{table}

\subsubsection*{FeCl$_2$} The largest Curie temperature is found for FeCl$_2$ in the MoS$_2$ crystal structure where we obtain $T_\mathrm{C}$ = 202 K. The main reason for the high value of $T_\mathrm{C}$ is the large magnetic moment of 4 $\mu_B$ per Fe atom and an exchange coupling of $J=15$ meV, which is one of the largest values found in the present study. Previous \textit{ab initio} calculations have reported that FeCl$_2$ in the CdI$_2$ crystal structure (which is metallic) is more stable compared to the MoS$_2$ crystal structure\cite{Torun2015, Feng2018} and the Curie temperature was estimated to 17 K based on mean-field theory.\cite{Torun2015} Our calculations confirm the stability hierarchy and predict an even more stable prototype GeS$_2$ (formation energy reduced by $\sim$ 30 meV/atom compared to the MoS$_2$ phase). However, we do not expect out-of-plane long-range ferromagnetic order in either the CdI$_2$ or the GeS$_2$ crystal structures, since the spinwave gaps are negative in both cases.
Interestingly, FeCl$_2$ in the CdI$_2$ crystal structure has positive single-ion anisotropy ($A$), which could indicate magnetic order. However, a negative anisotropic exchange coupling ($B$) yields an overall negative spinwave gap and the material thus serves as a good example of a case where the single-ion anisotropy is not a good indicator of magnetic order. To our knowledge there is no experimental reports of isolated 2D FeCl$_X$ compounds. However, FeCl$_3$ in the BiI$_3$ crystal structure has been intercalated in bulk graphite exhibiting a ferromagnetic transition at temperature $T = 8.5$ K.\cite{Dresselhaus2002}. Recently it has also been employed as functional intercalation in few-layer graphene compounds to weaken restacking of graphene sheets\cite{Qi2015} and bilayer graphene compounds\cite{Kim2011} to promote magnetic order in graphene.\cite{Bointon2014} Nevertheless, according to our calculations, the FeCl$_3$ crystal structure is less stable than the FeCl$_2$ ones and is not expected to exhibit ferromagnetic order as free standing layers due to a negative value of the spinwave gap. In bulk form FeCl$_2$ is known in the CdI$_2$ crystal structure with in-plane ferromagnetic order,\cite{McGuire2017} but the long range order is stabilized by interlayer anti-ferromagnetic exchange coupling, which supports our assertion that exfoliated layers of this type will not exhibit magnetic order. Bulk FeCl$_3$ has also been reported to form different stacking polymormphs of the BiI$_3$ crystal structure, but the magnetic properties of these materials are not known.\cite{McGuire2017}

\subsubsection*{MnO$_2$} Monolayers of MnO$_2$ in the CdI$_2$ crystal structure have been exfoliated in 2003,\cite{Omomo2003} but the magnetic properties have not yet been thoroughly investigated experimentally. Our calculations confirm a ferromagnetic ground in agreement with previous calculations\cite{Kan2013}, where a critical temperature of 140 K was predicted. However, that result were obtained from energy mapping analysis using PBE+U DFT calculations (U = 3.9 eV) and MC calculations based on the Ising model. From simulations of the Heisenberg model - explicitly including the finite anisotropy - we obtain a $T_\mathrm{C})$ of 63 K using Heisenberg parameters from a pure PBE calculations. The effect of Hubbard correction will be discussed in the next section.

\subsubsection*{YCl$_2$}A critical temperature of $T_C=55$ K is found for YCl$_2$ in the MoS$_2$ prototype. There is neither experimental or theoretical reports on this material and it could pose an interesting new 2D magnetic compound. Our calculations indicate that it is highly stable in the ferromagnetic configuration with magnetic moment of 1 $\mu_B$ per Y atom. However, since the material comprises a spin-1/2 system the classical MC calculations of the critical temperature may not be very accurate.

\subsubsection*{CrX$_3$}As reported in a previous study employing the same method\cite{Torelli2018} we predict CrI$_3$ in the BiI$_3$ structure to have $T_C=31$ K, while the similar compounds and CrCl$_3$ and CrBr$_3$ have $T_\mathrm{C}$ of 9 K and 19 K respectively. We note that our calculated critical temperature for CrI$_3$ is somewhat lower than the experimental value of 45 K. This is mainly due to the fact that PBE tends to underestimate the exchange coupling and can be improved by using a PBE+U scheme as discussed below. CrCl$_3$ and CrBr$_3$ have not previously been described in their 2D form, but are known as ferromagnetic bulk compounds consisting of layers in the BiI$_3$ crystal structure with out-of-plane magnetization.\cite{McGuire2017} The experimental Curie temperatures of bulk CrCl$_3$, CrBr$_3$, and CrI$_3$ are 27 K, 47 K, and 70 K respectively. Our calculated values show the same hierarchy, but are reduced compared to the bulk values due to the lack of stabilization by interlayer exchange coupling.

\subsubsection*{CuCl$_3$} For CuCl$_3$ in the BiI$_3$ crystal structure we find a critical temperature of 33 K, which is similar to the calculated value of CrI$_3$. The material does, however, lie above the convex hull by 0.15 eV per atom, which could complicate experimental synthesis and characterization.

\subsubsection*{XCl$_2$} Bulk CoCl$_2$ and NiCl$_2$ are both known to display anti-ferromagnetic interlayer coupling and in-plane ferromagnetic order \cite{McGuire2017}. As seen in Tab. \ref{tab:no_u}, our calculations predict the materials to exhibit out-of-plane order. For NiCl$_2$, however, one should be a bit cautious due to the extremely small value of the spinwave gap $\Delta$ and more accurate calculations could lead to a ground state with in-plane magnetic order. Experimental measurements on bulk samples indicate anomalies in the heat capacity related to magnetic phase transitions at 24 K and 52 K. While the first result is in good agreement with our predicted properties, the second one is significantly higher and could be related to an additional phase transition in the 3D structure.

\subsubsection*{Metastable high-$T_\mathrm{C}$ compounds} The lower part of Tab. \ref{tab:no_u} shows two materials that we do not predict to be stable, but may be of interest due to the large predicted critical temperatures. Here we comment briefly on the case of RuCl$_2$ in the MoS$_2$ crystal structure, which we predict to be a dynamically stable insulator with a critical temperature of 598K. It is, however, situated 0.5 eV above the convex hull, which will mostly likely pose an obstacle to experimental synthesis. Nevertheless, the calculations show that very high values of critical temperatures are indeed possible in 2D materials with realistic atomic-scale parameters. It should be mentioned that monolayers of RuCl$_3$ in the BiI$_3$ crystal structure have been exfoliated and characterized experimentally.\cite{Lippmann2016} Moreover, in a recent study\cite{Sarikurt2018} the critical temperature of monolayer RuCl$_3$ was calculated using DFT and MC simulations based on the Heisenberg model and found $T_C = 14.21$ K.\cite{Sarikurt2018} However, a Hubbard term is required to open a gap and RuCl$_3$ in the BiI$_3$ crystal structure is metallic within PBE,\cite{Huang2017,Ersan2018} which is why we do not include it in Tab. \ref{tab:no_u}.

\subsubsection*{In-plane anisotropy} As mentioned above, materials with the TMDH crystal structure have been considered as effective triangular magnetic lattices with a single nearest-neighbour coupling. However, this model can only be used for a rough screening of materials. For example, CrIS exhibits a strong in-plane anisotropy and the axis of magnetization are ordered (from the hardest axis to the easiest) as: $x$, $z$ and $y$. In Eq. \eqref{eq:heis}, in-plane anisotropy is not considered and we thus extend the model with the full set of anisotropy parameters $A_x$, $A_y$, $B_x$, and $B_y$ that measures the single-ion anisotropy and anisotropic exchange with respect to both $x$ and $y$ directions (relative to the $z$-direction). These parameters can be found by generalizing the energy mapping analysis Eqs. \eqref{eq:A}-\eqref{eq:B} to include different in-plane directions. We then run MC calculations using the full set of parameters to find the critical temperatures, including nearest and next-nearest neighbours couplings $J_1$ and $J_2$.
We find three insulating materials in this crystal structure that shows ferromagnetic order. The results are shown in Tab. \ref{tab:maganis_feocl}. In particular, CrIS is predicted to have a critical temperature of 140 K. We also note that we obtain a Curie temperature of 15 K for CrClO, which has previously been predicted to have a Curie temperature of 160 K based on an Ising model approach.\cite{Miao2018} Again, this comparison emphasizes that the magnetic anisotropy cannot simply be regarded as a mechanism that fixes the magnetization to the out-of-plane direction: approximating magnetic properties by the Ising model may yield a critical temperature that is wrong by an order of magnitude.
\begin{table}[tb]
  \begin{center}
\begin{tabular}{lcccccccc}
        & $J_1$ & $J_2$ & $A_x$ &  $A_y$ & $B_x$ & $B_y$ & Easy axis & $T_\mathrm{C}$      \\
    \hline
CrIS       & 5.71 & 4.85 &    0.084  &  -0.223  &  0.025  &  0.033   & x,z,\textbf{y} &  140 \\
MnClN       &  2.66 & 5.76 & 0.023  &  0.044  &  0.022  &  0.012   & x,y,\textbf{z} &  75 \\
CrClO       & 1.08 & 0.74 &  -0.010  &  0.034  &  0.004  &  0.001    & y,x,\textbf{z}  &  15\\

\end{tabular}
\end{center}
\caption{Ferromagnetic materials in the TMDH crystal structure. The first two columns show nearest neighbour and next-nearest neighbour exchange coupling constants in meV. Columns three to six display anisotropy parameters calculated with respect to the two in-plane directions $x$ and $y$ in meV. In the second last column we state the crystallographic directions of magnetization listed from the hardest to the easiest axis. The last column shows the critical temperature in K obtained from MC calculations with these parameters.\label{tab:maganis_feocl}}
\end{table}

\subsection{Hubbard U}
\begin{figure}[tb]
	\includegraphics[width = 0.5\textwidth]{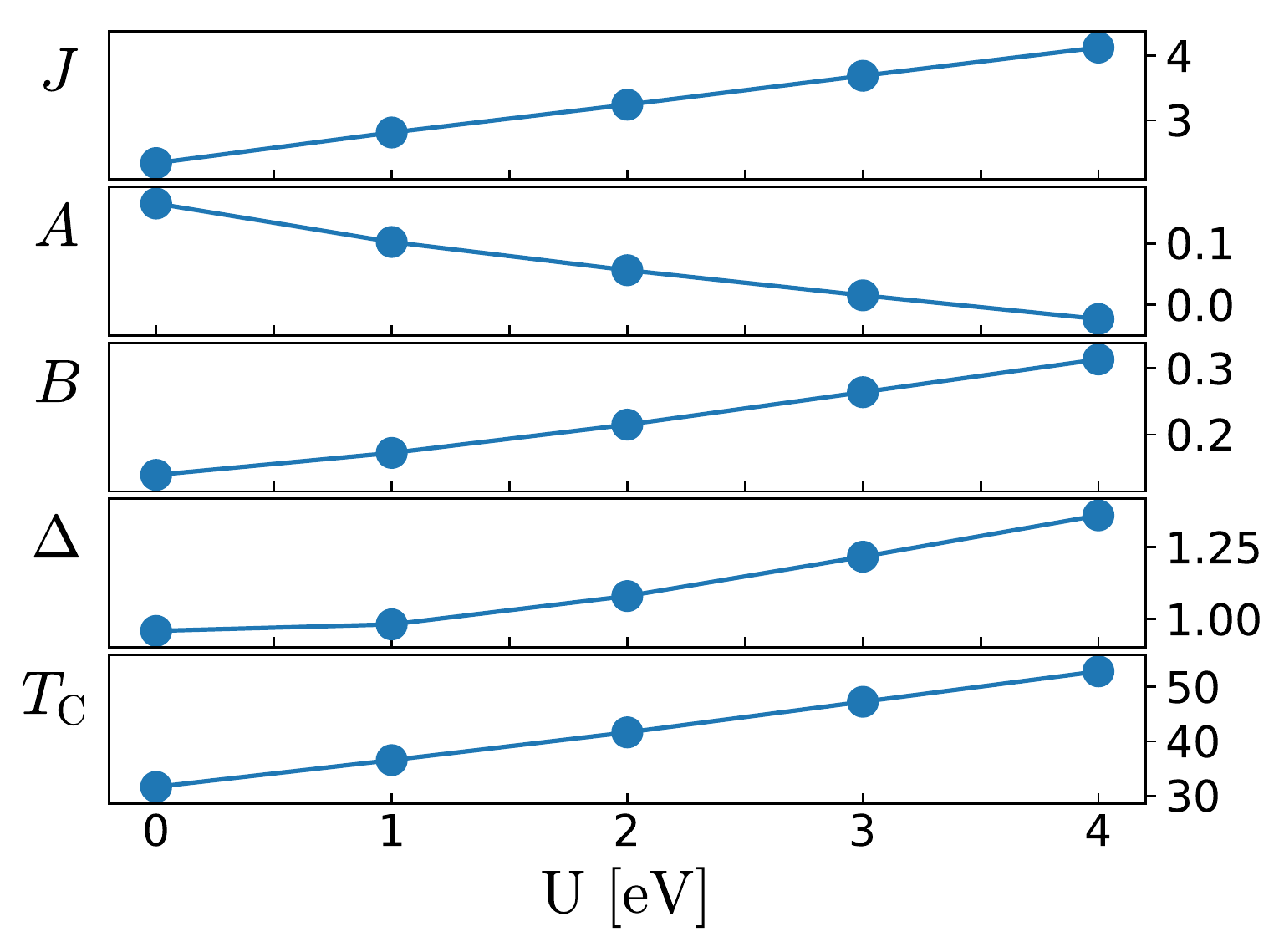}
	\caption{Calculated magnetic parameters of CrI$_3$ as a function of U. $J$, $A$ and $B$, $\Delta$ are in units of meV and critical temperatures $T_{C}$ are in units of K.\label{fig:cri3}}
\end{figure}

Almost half of the materials present in C2DB contain at least one element with a partially filled $d$-shell. Local and semi-local xc-functionals such as PBE are known to overestimate delocalization of correlated electrons, due to the uncompensated Coulomb self-interaction of the electron. In the Hubbard model a term is introduced that acts as an effective electronic on-site repulsion and provides a penalty to delocalization. In order to determine the influence of the Hubbard correction we have recalculated exchange and anisotropy parameters for CrI$_3$ for a range of U values in the PBE+U scheme. The structure was fully relaxed for each value of U and the results are shown in Fig. \ref{fig:cri3}. We observe that an increasing value of U leads to an overall increase of both $\Delta$ and $J$, which result in higher critical temperatures. The dependence of $T_\mathrm{C}$ on U is roughly linear with $T_\mathrm{C}$ increasing by 5 K per eV that U is increased.

In order to gain more insight into the general influence of U for the calculations of magnetic properties, we have recalculated the magnetic parameters and critical temperature for all magnetic materials in the C2DB containing 3$d$ valence electrons. We used the optimal values determined in Ref. \onlinecite{Wang2006} and listed in Tab. \ref{tab:u_value}. For each material the structure was relaxed with the given value of U, but the stability analysis was based on bare PBE.
\begin{table}[tb]
  \begin{center}
\begin{tabular}{l|c|c|c|c|c|c|c}
    Element & Fe & Mn & Cr & Co & Ni & V & Cu \\
  \hline
    U [eV]        & 4.0 & 3.8 & 3.5 & 3.3 & 6.4 & 3.1 & 4.0  \\
\end{tabular}
\end{center}
\caption{Hubbard parameters employed in the PBE+U calculations.\label{tab:u_value}}
\end{table}

\begin{figure*}[tb]
	\includegraphics[width = \textwidth]{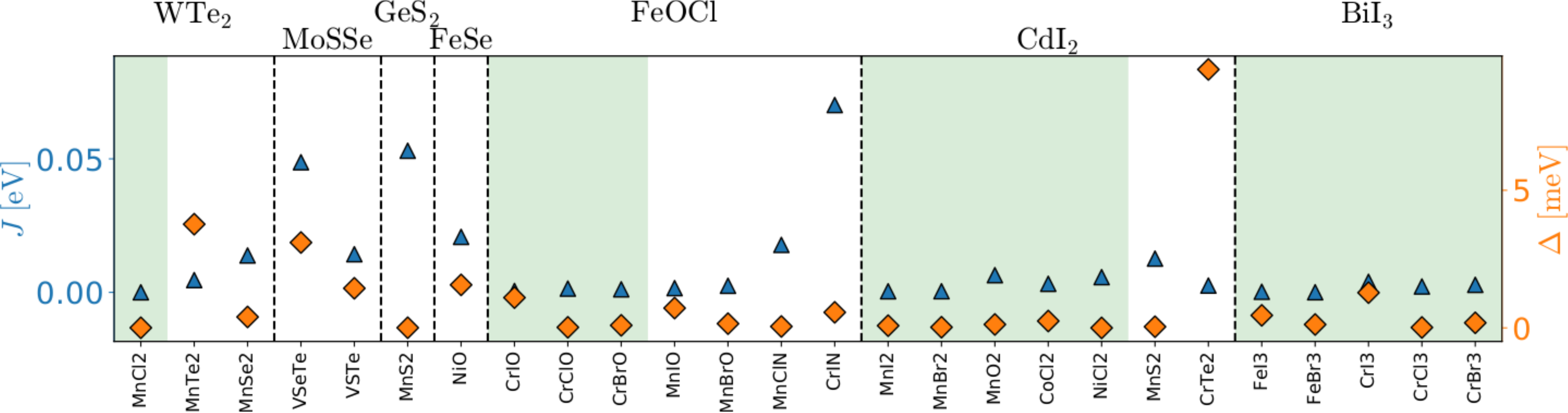}
	\caption{PBE+U calculations of exchange coupling $J$ (triangles) and spinwave gap $\Delta$ (squares) of stable ferromagnetic materials with $\Delta >0$. Green background indicates insulating materials. \label{fig:tc_u}}
\end{figure*}
The inclusion of U can have a rather dramatic effect on the results. For example, the magnetic configuration of the ground state or the magnetic moment localized on the transition metal ion may change. The results for the stable materials are shown in Fig. \ref{fig:tc_u}, while the insulating systems are listed in Tab. \ref{tab:with_u}. Including a Hubbard term in the entire workflow (from the relaxation step onward) affects quantitatively and in some cases also qualitatively the ground-state. This means that the magnetic moment, the energy gap or the sign of $\Delta$ may change, making the comparison with Tab. \ref{tab:no_u} meaningful only for a subset of materials.  
Compounds that are also present in Tab. \ref{tab:no_u} are shown in bold face for comparison. 
\begin{table}[tb]
  \begin{center}
\begin{tabular}{lcccccc}
Formula & Structure & $J$ [meV] & $\Delta$ [meV] & $S$ [$\hbar$]& $T_\mathrm{C}$ [K]  \\
  \hline
  \hline
\textbf{MnO$_2$}  &   CdI$_2$  &   6.43 &  0.125  &  1.5  &  82 \\
\textbf{CoCl$_2$}  &   CdI$_2$  &   3.21 &  0.249  &  1.5  &  57  \\
\textbf{CrI$_3$}  &   BiI$_3$  &   3.95  &  1.280  &   1.5  &  50  \\
\textbf{CrBr$_3$}  &   BiI$_3$  &   2.82  &  0.185   &  1.5  &  24  \\
MnI$_2$  &   CdI$_2$  &   0.40 &  0.081  &   2.5  &  21  \\
MnBr$_2$  &   CdI$_2$  &   0.41 &  0.024  &   2.5  &  16 \\
\textbf{CrCl$_3$}  &   BiI$_3$  &   2.19  &  0.016   &  1.5  &  10  \\
\textbf{NiCl$_2$}  &   CdI$_2$  &   5.69 &  $\sim10^{-4}$  &    1.0  &  7 \\
FeBr$_3$  &   BiI$_3$  &   0.04  &  0.124  &  2.5  &  2  \\
\hline
CoO  &   FeSe  &   106.54  &  0.199  &  1.5  &  520  \\
FeS  &   FeSe  &   28.99  &  0.591  &   2.0  &  413  \\
\end{tabular}
\end{center}
\caption{List of 2D non-metal materials with positive exchange coupling $J$ and spin wave gap $\Delta$, obtained from PBE+U calculations. Structure denotes the prototypical crystal structure and $S$ is the spin carried by each magnetic atom. The structures in bold are present also in \ref{tab:no_u} for comparison. The top part contains stable materials, whereas the lower part contains materials with large critical temperatures that may be unstable in their pristine form. \label{tab:with_u}}
\end{table}

\subsubsection*{Effect of U on $T_\mathrm{C}$}
For MnO$_2$, the main effect of adding a Hubbard correction is to increase the exchange parameter $J$ by a factor of two. Interestingly the anisotropy parameters $A$ and $B$ decrease and the spinwave gap  $\Delta$ becomes less than half the value obtained with PBE. Nevertheless, the overall effect is an increase of the critical temperature from 63 K to 82 K. This number can be compared to the result in Ref. \onlinecite{Kan2013} where a critical temperature of 140 K was estimated from the Ising model.

For CrI$_3$ we predict a critical temperature of 50 K with PBE+U. This is significantly closer to the experimental value than the 31 K obtained with bare PBE. Similarly $T_\mathrm{C}$ increases from 19 K to 24 K for CrBr$_3$ while the $T_\mathrm{C}$ of CrCl$_3$ is increased from 9 K to 10 K. The critical temperature of CoCl$_2$ is almost unaffected. These results indicate that it is non-trivial to predict how the inclusion of a Hubbard U influences the calculated critical temperatures in general.

For the compounds MnI$_2$, MnBr$_2$, and FeBr$_3$, which all have large magnetic moment of $5\mu_\mathrm{B}$ per magnetic atom we obtain rather low critical temperatures of 21, 16 and 2 K respectively. This is mainly due to small values of exchange coupling $J$ for these materials. The inclusion of U in MnI$_2$ and  MnBr$_2$ increases the electronic gap as well as the spinwave gap. But most importantly, it yields a ferromagnetic ground state, while the ground state is anti-ferromagnetic without the inclusion of U.\cite{Kulish2017} For MnI$_2$ The result appears to be in qualitative agreement with neutron scattering experiments on the bulk compounds, which reports a helical magnetic structure below a critical temperature of 3.4 K, with the moments being aligned in the individual planes.\cite{Cable1962, McGuire2017} This could indicate that PBE+U provides a more accurate description than PBE, which does not predict magnetic order for MnI$_2$. For MnBr$_2$ in the CdI$_2$ crystal structure, neutron scattering experiments on the bulk parent structure revealed an anti-ferromagnetic order below $T=2.16$ K with magnetic moments lying in-plane.\cite{Wollan1958} However, this is not necessarily in contradiction with our calculations since the observed anti-ferromagnetic configuration is "double-striped", a configuration that has not been considered in the present study. For FeBr$_3$ the Hubbard term makes the spin jump from $\hbar/2$ to $5\hbar/2$per Fe atom and opens a spinwave gap. A previous investigation of this material showed that it is predicted to be a quantum spin Hall insulator with PBE while PBE+U predicts a trivial insulator above a critical value of U$=0.18$ eV.\cite{Olsen}

\subsubsection*{Metastable high-$T_\mathrm{C}$ compounds} The lower part of Tab. \ref{tab:with_u} lists materials, which are not predicted to be completely stable in their pristine form according to PBE calculations (we have not performed a full stability analysis with PBE+U). Bulk CoO has an anti-ferromagnetic rock-salt structure with a critical temperature of 293 K\cite{Berkowitz1999}. According to PBE calculations the most stable 2D phase is a metallic CdI$_2$ crystal structure (parameters $J$ and $\Delta$ are shown in Fig. \ref{fig:tc_nou}). In the FeSe crystal structure, CoO has a low dynamic stability but we report it here due to the very high critical temperature of 520 K originating from the extraordinarily large exchange coupling predicted by PBE+U. 

FeS in the FeSe crystal structure has a non-magnetic ground state with PBE, but is predicted to be highly stable and is situated on the convex hull. With PBE+U the ground state becomes ferromagnetic and we predict a high critical temperature of 413 K. According to previous calculations,\cite{Winiarski2018} however, the true ground state is a striped anti-ferromagnetic configuration, which is not taken into account in this work.

\begin{table}[tb]
  \begin{center}
\begin{tabular}{lcccccccc}
        & $J_1$ & $J_2$ & $A_x$ &  $A_y$ & $B_x$ & $B_y$ & Easy axis & $T_\mathrm{C}$      \\
    \hline
   CrBrO    & 1.12 & 0.78  &  0.043  &  -0.010  &  0.001  &  0.001 &  x,z,\textbf{y} &  35 \\
CrIO     & 0.49 & -1.46 & 0.586  &  -0.123  &  -0.008  &  -0.003& x,z,\textbf{y}  &  25 \\
    \textbf{CrClO}   & 1.38 & 1.27 & 0.007  &  0.016  &  0.001  &   $\sim-10^{-4}$  & y,x,\textbf{z} &  20\\

\end{tabular}
\end{center}
\caption{  Parameters and results for TMHC structures obtained from PBE+U calculations. Symbols and units are the same as in Tab.  \ref{tab:maganis_feocl}. Materials in bold are the ones listed in both tables. \label{tab:maganis_feocl_u}}
\end{table}

\subsubsection*{In-plane anisotropy}
In Tab. \ref{tab:maganis_feocl_u} we list Heisenberg parameters and critical temperatures for TMHC structures obtained from PBE+U calculations and MC calculations following the same procedure as in the previous section where no Hubbard correction was included. Comparing the results with Tab. \ref{tab:maganis_feocl}, it is noted that MnClN and CrIS are not predicted to be ferromagnetic insulators with PBE+U. In particular, MnClN is predicted to be a metal and CrIS exhibits a negative spinwave gap. On the other hand, two new materials - CrIO and CrBrO - are predicted to exhibit ferromagnetic order at 25 K and 35 K respectively.

\section{Conclusions}\label{sec:conclusions}
We have presented a high throughput computational screening for magnetic insulators based on the Computational 2D Materials Database. In contrast to several previous studies of magnetism in 2D, we have emphasized the crucial role of magnetic anisotropy and used the spinwave gap as a basic descriptor that must necessarily be positive in order for magnetism to persist at finite temperatures. This criterion severely reduces the number of relevant candidates and we end up with 12 stable candidate materials for which the critical temperatures were calculated from classical MC simulations. Seven of the materials were predicted to have Curie temperatures exceeding that of CrI$_3$.

The classical MC simulations appear to comprise an accurate method for obtaining the critical temperatures for insulating materials with $S>1/2$. However, the Heisenberg parameters that enter the simulations may be sensitive to the approximations used to calculate them. In the C2DB all calculations are performed with the PBE functional, which may have shortcomings for strongly correlated systems. We have thus tested how the results are modified if the parameters are evaluated with PBE+U instead and we find that the predictions do indeed change in a non-systematic way. For the hexagonal and honeycomb systems three materials that were predicted to be ferromagnetic (at finite temperature) are no longer predicted to show magnetic order when the PBE+U scheme is employed and three materials that were not magnetic with PBE become magnetic with PBE+U. For the five materials that are magnetic with both PBE and PBE+U the critical temperatures are slightly different in the two approximations. The biggest difference is seen for CrI$_3$ where inclusion of U increases the critical temperature from 31 K to 50 K, which is closer to the experimental value of 45 K.

In the present work we have mainly focused on insulators. This restriction is rooted in the simple fact that we do not have a reliable way to estimate Curie temperatures of metallic 2D magnetic materials. Metallic ferromagnetism in 2D is, however, a highly interesting subject and we note that room-temperature magnetism has recently been reported in the 2D metals VSe$_2$\cite{Bonilla2018} and MnSe$_2$.\cite{OHara2018} Moreover, Figs. \ref{fig:delta_j} and \ref{fig:tc_nou} indicate that in the C2DB the largest values of both spinwave gaps and exchange couplings are found in metallic materials. Clearly, there is pressing need for theoretical developments of 2D itinerant magnetism that can be applied in conjunction with first principles simulations to provide accurate predictions of the magnetic properties of 2D metallic materials.

Finally, we have restricted ourselves to ferromagnetic order. Nevertheless, the C2DB contains 241 anti-ferromagnetic entries - 50 of which are predicted to be stable. The prediction of a novel 2D anti-ferromagnetic compound would certainly comprise an important step forward in the study of 2D magnetism, but the theoretical treatment is complicated by the possibility of several ordered structures that may coexist at a given temperature - in particular for non-bipartite lattice such as the triangular one. We will leave the study of anti-ferromagnetism in 2D to future work.

\section{Acknowledgements}
 T. Olsen and D. Torelli acknowledges supported from the Independent Research Fund Denmark, grant number 6108-00464A. K. Thygesen acknowledges support from the European Research Council (ERC) under the European Union’s Horizon 2020 research and innovation program (Grant No. 773122, LIMA).


%

\end{document}